\documentclass[letterpaper, 10 pt, conference]{ieeeconf}  

\IEEEoverridecommandlockouts                              

\usepackage{amssymb,amsthm,amsmath}
\usepackage{comment}
\usepackage{graphicx}
\usepackage{algorithm}
\usepackage{moresize}
\usepackage{multirow}
\usepackage{multicol}
\usepackage{algpseudocode}

\usepackage{tikz}
\usepackage{cite}

\theoremstyle{definition}
\newtheorem{definition}{Definition}
\newtheorem{remark}{Remark}

\DeclareMathOperator*{\argmax}{arg\,max}

\title{\LARGE \bf
    Capability Augmentation for Heterogeneous Dynamic Teaming \\ with Temporal Logic Tasks}

\author{Carter Berlind, Wenliang Liu, Alyssa Pierson, and Calin Belta
\thanks{This work was partially funded by the NSF under grant IIS-2024606 and a Boston University startup grant}
\thanks{The authors are with the Department of Mechanical Engineering, Boston University, Boston, MA, USA {\tt\small cberlind@bu.edu, wliu97@bu.edu, pierson@bu.edu, cbelta@bu.edu}}}%
        
\begin{document}

\maketitle

\begin{abstract}
This paper considers how heterogeneous multi-agent teams can leverage their different capabilities to mutually improve individual agent performance. We present Capability-Augmenting Tasks (CATs), which encode how agents can augment their capabilities based on interactions with other teammates. Our framework integrates CAT into the semantics of Metric Temporal Logic (MTL), which defines individual spatio-temporal tasks for all agents. A centralized Mixed-Integer Program (MIP) is used to synthesize trajectories for all agents. We compare the expressivity of our approach to a baseline of Capability Temporal Logic Plus (CaTL+). Case studies demonstrate that our approach allows for simpler specifications and improves individual performance when agents leverage the capabilities of their teammates. 
\end{abstract}

\section{Introduction}
\label{sec:intro}

In this paper, we consider a team of agents with varying capabilities and individual tasks, i.e. each agent has a task that it alone is responsible for completing. We focus on tasks where agents need to traverse an environment to reach regions where they can fulfill service requests while avoiding dangerous regions. In certain cases, we identify that agents may be able to fulfill requests without being in the proper regions and stay safe in dangerous regions if they are accompanied by agents with complementary capabilities. 

For example, consider a ground agent that must stay dry but travel through regions with water if it is carried by an aerial agent. Collaboration between these agents could improve their ability to satisfy their tasks. We call this {\em capability-augmenting collaboration}.

In this work, we focus on generating high-level motion plans for a team of agents using capability-augmenting collaboration. We assign the agents spatio-temporal tasks that include servicing requests and avoiding danger using a rich specification language and formulate a planner to generate trajectories for the team of agents. Our goal is to maximize the number of agents that complete their tasks and minimize the distance they travel.


Previous works explore agents satisfying tasks in heterogeneous teams \cite{SanjaySarma2020,Kim2022,Ramachandran2019,Twu2014,Prorok2017}. \cite{SanjaySarma2020} demonstrates improvements in rescues made in search and rescue settings, \cite{Kim2022} shows improvements in coverage of domains where regions require different mobility capabilities, and \cite{Ramachandran2019} shows improvements resilience in networked systems. These works provide ample justification for further exploitation of heterogeneity in multi-agent systems. Works such as \cite{Notomista2022,Miyano2023,Miyano2022,Jia2013,Dias2006,Berman2009,Iijima2017,Antonyshyn2023} provide a foundation for how various tasks are allocated and completed using a team of robots based on agent capabilities. The authors of \cite{Notomista2022} assign tasks based on system resilience and energy efficiency use, and \cite{Miyano2023} and \cite{Miyano2022} assign tasks based on cooperation constraints between agents. In this paper, we build on these works by formalizing a method for agents with tasks that are already allocated to better leverage their heterogeneity through capability-augmenting collaboration.


We use Metric Temporal Logic (MTL) \cite{koymans1990specifying}\cite{FAINEKOS20094262} to formalize the spatio-temporal tasks for the agents. MTL is a rich specification language that allows us to assign tasks based on concrete time intervals.
Works such as \cite{Liu2023,Cardona2022,Leahy2022,Buyukkocak2021} offer task allocation and planning formulations for heterogeneous teams using temporal logic specifications. For instance, the authors of \cite{Liu2023} defined specifications with Capability Temporal Logic plus (CATL+) in a disaster response scenario, and \cite{Cardona2022} defined a construction task for modular and reconfigurable robots using MTL. It is often difficult or intractable to specify that tasks can be satisfied using capability-augmenting collaboration in these formulations. We provide a detailed comparison between previous methods and ours in Section \ref{sect:expressivity} to demonstrate our method's benefits. 

In this paper we build on the tools designed for specification and verification of behaviors of teams of robots, such as \cite{Leahy2022,Leahy2022a,Sun2022,Cardona2022,Schillinger2018,Cai2022,Luo2022,Ulusoy2012,Kantaros2019,Bai2022}, by introducing a formalization for capability-augmenting collaboration using MTL. We use a special case of MTL in which we introduce Capability-Augmenting Tasks (CATs) to simplify the encoding of capability-augmenting collaboration.
Specifically, the contributions of this work are as follows: 
\begin{enumerate}
    \item We formalize a planning problem for a team of heterogeneous agents using capability-augmenting collaboration in MTL.
    \item We formulate a centralized Mixed Integer Program (MIP) to generate trajectories for the agents.
    \item We demonstrate the benefit of capability-augmenting collaboration in satisfying individual tasks in comparison to systems where it is not used.
\end{enumerate} 

We formalize the problem of agents utilizing capability-augmenting collaboration and motivate the need to generate group trajectories in Sec.~\ref{Sect:Problem-Statement}. We present an approach for generating motion plans that satisfy as many individual tasks as possible while minimizing the distance the agents travel using an MIP in Sec.~\ref{Sect:Approach}. In Sec.~\ref{sect:expressivity}, we compare the types of collaboration that our formulation can capture to previous work. Finally, we use a case study to demonstrate the benefits of using capability-augmenting collaboration in Sec.~\ref{Sect:Case-Studies}. 





\section{Problem Statement}
\label{Sect:Problem-Statement}
In this section, we formalize the planning problem for a team of agents using capability-augmenting collaboration to satisfy individual tasks. First, we provide a model for the agents and environment. We then use this model to define individual tasks for the agents using metric temporal logic. We then introduce our method for encoding capability-augmenting collaboration using CATs. Finally, we use our model of the agents and environment, along with the MTL tasks, to formalize a planning problem that we can approach solving in the next section.

To create this formalization, we introduce some basic notation. In this work, we use $\mathbb{R}_{\geq0}$ to describe the set of non-negative real numbers, and $\mathbb{Z}_{\geq 0}$ denotes the set of non-negative integers. $|\cdot|$ denotes the cardinality of a set. $||\cdot||_1$ denotes the L-1 norm. Let $2^{S}$ denote the power set of set $S$.



\subsection{System Model}
\label{Subsect:Model-Notation}

In this section, we formalize a model for the agent and environment. We consider a group of heterogeneous agents operating in a partitioned 2D shared environment. In each region, there can either be a service request that an agent can satisfy by visiting that region, an indication that the region should be avoided, or neither.
Formally, we define the environment as a tuple $Env = (\mathcal{Q},E,\Pi,\mathcal{L})$, where $\mathcal{Q}$ is the set of nodes (vertices) corresponding to environment regions and $E\subseteq \mathcal{Q}\times \mathcal{Q}$ is the set of directed edges such that  $(q,q')\in E$ iff an agent can transition from node $q$ to node $q'$
(this includes self-transitions). The environment we consider is a grid such that all transitions that are not self-transition are the same distance. For $q\in \mathcal{Q}$, let $adj(q)\subset \mathcal{Q}$ be the set of nodes that are adjacent to node $q$, i.e., $q'\in adj(q)\iff (q,q')\in E$.  Let $\Pi$ be a set of labels that report the the service requests and indications of danger of the environment, and $\mathcal{L}: \mathcal{Q}\mapsto 2^\Pi$ 
be a mapping that matches each node to a set of labels. Agents are asked to visit and avoid the features of the environment by referencing if it is in a node with a corresponding label.

We assume that the agents travel between nodes using existing edges. As they travel, they incur costs based on the number of transitions they take that are not self-transitions. The agents are synchronized using a central clock. At every time step the agents either transition to a new node or make a self-transition and stay in place. These transitions always take a single time step. 
Agents are indexed from a finite set $\{1,...,N_A\}$, where $N_A$ is the total number of agents in the environment.

We define an individual agent as a tuple $a_j=(\mathcal{C}_j,q_j(0))$, where $\mathcal{C}_j$ is the set of capabilities of agent $j$, $j\in\{1,...,N_A\}$. The capabilities of each agent determine how they collaborate with other agents (detailed in Sec.~\ref{subsec:MTL} and Sec.~\ref{subsec:CAT}). Let $\mathcal{C}_{global}$ be the total set of capabilities held by all the agents in the system, i.e., $\mathcal{C}_{global}=\cup_{j\in\{1,\ldots,N_A\}}\mathcal{C}_j$. We define $I_c$ as the set of indices of agents with a specific capability $c$ where  $c\in \mathcal{C}_{global}$, i.e. $j\in I_c\iff c\in\mathcal{C}_j$. Let $q_j(k)\in\mathcal{Q}$ be the node occupied by agent $j$ at time $k$, and $q_j(0)$ be the agent's initial node. In this paper, we only consider the high-level motion plan, i.e., the transitions of robots between nodes. We assume any number of agents can be in the same node and take the same transitions simultaneously without worry of collision. The inter-agent collision can be avoided by using a lower-level controller tracking the high-level motion plan, which will be considered in future work. 
We define a group trajectory $\mathbf{Q}\in\mathcal{Q}^{T_p\times N_A}$ as the sequence of nodes every agent visits where $\mathbf{Q}[k]=[q_1(k),\ldots,q_{N_A}(k))]$ and $T_p$ is the planning horizon (defined in Sec.~\ref{subsec:problem}).

\subsection{MTL Specifications}
\label{subsec:MTL}
Next, we define our language for specifying tasks for the agents. 
In this paper, we use MTL \cite{koymans1990specifying}\cite{FAINEKOS20094262} to formally define an individual specification for each agent. We define the individual MTL specification for an agent $j$ over its \emph{trace}. The \emph{trace} of agent $j$ is defined as $\mathbf{O}_j=\mathbf{o}_j(0),\mathbf{o}_j(1),...,\mathbf{o}_j(T_p)$, where $\mathbf o_j(k)$ is the observation of agent $j$ at time step $k$, defined as $$\mathbf{o}_j(k) = (\mathcal{L}(q_j(k)),\{n^c_j(k)\}_{c\in\mathcal{C}_{global}}).$$ Here, $n_j^c(k)$ is the number of agents with capability $c$ in the same node with agent $j$ at time step $k$ excluding agent $j$: $$n_j^c(k) = \sum_{i\in I_c\setminus j} \beta\big(q_i(k)=q_j(k)\big),$$ where $\beta$ is a binary indicator function that returns $1$ if a statement is true and $0$ otherwise. The reason we include $n_j^c(k)$ is so we can evaluate tasks using capability-augmenting collaboration. The syntax of the MTL we use in this paper is defined as follows:
\begin{align}\label{Eqn:MTL-def}
    \phi :=\top|\text{AP}|\neg \phi|&\phi_1\land\phi_2|\phi_1\lor\phi_2|\notag\\
    &\lozenge_{[k_1,k_2]}\phi|\square_{[k_1,k_2]}\phi|\phi_1 U_{[k_1,k_2]}\phi_2.
\end{align}
Here, $k_1,k_2\in \mathbb{Z}_{\geq0}$ are time bounds with $k_1\leq k_2< \infty$. $\lozenge_{[k_1,k_2]}$ is the temporal operator {\em eventually}, and $\lozenge_{[k_1,k_2]}\phi$ requires that $\phi$ has to be true at some point in the time interval $[k_1,k_2]$. $\square_{[k_1,k_2]}$ is the temporal operator {\em always}, and 
$\square_{[k_1,k_2]}\phi$ requires that $\phi$ must be true at all points in the time interval $[k_1,k_2]$. $U_{[k_1,k_2]}$ is the temporal operator {\em until}, and $\phi_1 U_{[k_1,k_2]}\phi_2$ requires that $\phi_2$ is true at some point in $[k_1,k_2]$ and $\phi_1$ is always true before that. $\neg$, $\land$, and $\lor$ are the negation, conjunction, and disjunction Boolean operators, respectively. AP is an atomic proposition. We will formally encode capability-augmenting collaboration into the atomic propositions in Section \ref{subsec:CAT}. We use $(\mathbf O_j,k)\models\phi$ to denote that the trace of agent $j$ satisfies the MTL formula $\phi$ at time $k$. The time horizon of a formula $\phi$, denoted as $T_{\phi}$, is the smallest time point in the future for which the observation is required to determine the satisfaction of the formula. Formal definitions of the semantics, as well as the time horizon $T_{\phi}$ can be found in \cite{koymans1990specifying} and \cite{FAINEKOS20094262}.  

\subsection{Capability-Augmenting Task}
\label{subsec:CAT}
In this subsection, we define a specific kind of atomic proposition called CATs that we use to formalize and encode capability-augmenting collaboration into MTL.

\begin{definition}[Syntax of CATs]
    The syntax of a CAT is defined as:
\begin{align}\label{Eqn:CAT-Definition}
    \mathcal{CAT} =\langle \pi,\{c_{aug},m_{aug}\},\{c_{al},m_{al}\}\rangle,
\end{align}
where $\pi\in\Pi$ is a label, $c_{aug}, c_{al}\in \mathcal C_{global}$ are two capabilities, $m_{aug}, m_{al}\in \mathbb Z_{\geq 0}$ are two non-negative integers.
\end{definition}

Intuitively, an agent $j$ can satisfy the CAT \eqref{Eqn:CAT-Definition} at time $k$ in two ways. The first is to reach a node labeled by a specified $\pi$, i.e., $\pi\in\mathcal L(q_j(k))$. In cases when agent $j$ cannot reach a node with label $\pi$, the second way that the agent may be able to satisfy the CAT is that there are at least $m_{aug}$ agents with a compatible capability $c_{aug}\in \mathcal{C}_{global}$ and less than $m_{al}$ agents with capability $c_{al}\in \mathcal{C}_{global}$ in the agent's node $q_j(k)$ excluding itself. We refer to $c_{aug}$ as an augmenting capability and $c_{al}$ as an availability-limiting capability.
In other words, if the required number of agents with an augmenting capability $c_{aug}$ are present and available in an agent's node, then the agent can satisfy the CAT with their assistance without being in a node with the required label $\pi$. The agents with the augmenting capability are unavailable if and only if there are no less than $m_{al}$ agents with the availability-limiting capability $c_{al}$ in the agent's node excluding itself because we assume that the agents with the augmenting capability $c_{aug}$ may be assisting other agents. 
We make this conservative assumption to ensure that generating trajectories based on CATs is tractable.
We will address relaxing this assumption in future work. The second way to satisfy a CAT, i.e., satisfying it via collaboration, is referred to as \emph{capability-augmenting collaboration}. The satisfaction of a CAT, referred to as the qualitative semantics, is formulated in the following definition. 

\begin{definition}[Qualitative Semantics of CAT]
The trace of agent $j$ satisfies a $\mathcal{CAT}=\langle \pi,\{c_{aug},m_{aug}\},\{c_{al},m_{al}\}\rangle$ at time $k$, i.e., $(\mathbf O_j,k)\models \langle \pi,\{c_{aug},m_{aug}\},\{c_{al},m_{al}\}\rangle$, if and only if:
\begin{equation}
\label{Eqn:CEP-Evaluation}
    \pi\in\mathcal{L}(q_j(k))\lor(n_j^{c_{aug}}(k)\geq m_{aug}
    \land n_j^{c_{al}}(k)<m_{al}).
\end{equation}
\end{definition}

There are two special cases for a CAT. First, if $m_{al}>|I_{c_{al}}|$, which means that agents with capability $c_{aug}$ are always available to collaborate, we remove $c_{al},m_{al}$ from the CAT for simplicity, i.e., the CAT is denoted as $\langle \pi,\{c_{aug},m_{aug}\},\{\}\rangle$. Second, if $m_{aug}>|I_{c_{aug}}|$, which means that agents with capability $c_{aug}$ can never collaborate with agent $j$, we remove $c_{aug},m_{aug}$ from the CAT and turn it into $\langle \pi,\{\},\{\}\rangle$ for simplicity. In both cases, the semantics \eqref{Eqn:CEP-Evaluation} can be evaluated without the omitted components.

\subsection{Problem Formulation}\label{subsec:problem}
We assign an MTL specification $\phi_j$ defined over $\mathbf O_j$ to each agent $j$. 
We define $\rho(\mathbf{O}_j,\phi_j,k)$ to quantify the satisfaction of $\phi_j$, where $\rho$ maps a trace $\mathbf{O}_j$, an MTL formula $\phi_j$, and time step $k$ to either $1$ or $-1$ depending on satisfaction. It is defined formally as follows:
\begin{align}
    \rho(\mathbf{O}_j,\phi_j,k)=&
    \left\{
        \begin{array}{lr}
             1,&\text{if }\mathbf{O}_j[k]\models\phi_j,\\
             -1,&\text{if }\mathbf{O}_j[k]\models\phi_j.
        \end{array}
    \right.
    \label{Eqn:Rho-def}
\end{align}

This metric does not report the degree of satisfaction as a robustness metric would. Rather, it numerically captures Boolean satisfaction for individual agents. We determine the planning horizon from the individual specifications as $T_p=\max_{j\in \{1,...,N_A\}}(T_{\phi_j})$.
We evaluate the performance of agent $j$ using $\rho(\mathbf{O}_j,\phi_j,0)$ and an individual motion cost. 
The motion cost  $J_j:\mathcal{Q}^{T_p\times N_A}
\mapsto\mathbb{Z}_{\geq 0}$ is the number of transitions, excluding self-transitions, an agent takes and is formulated as:
\begin{align}\label{Eqn:individual-cost}
    J_j&(\mathbf{Q})=\\
    &\sum_{k=0}^{T_p-1}\sum_{q\in\mathcal{Q}}\sum_{q'\in adj(q)\backslash q}\beta\big((q,q')=(q_j(k),q_j(k+1))\big).\notag
\end{align}
We define individual agent performance $P_j:\mathcal{Q}^{T_p\times N_A}\times\phi_j\mapsto\mathbb{R}$ as: 
\begin{equation}\label{eqn:ind-task-perf}
  P_j(\mathbf{Q},\phi_j) =  M*\rho(\mathbf{O}_j,\phi_j,0)-J_j(\mathbf{Q}),
\end{equation}
where $M$ is a large enough positive number $M\in \mathbb{R}_{\geq 0}$ to prioritize the satisfaction of $\phi_j$. Note that we use the nodes that makeup $\mathbf{Q}$ to solve for the sequence of observations $\mathbf{O}_j$, meaning that $\mathbf{Q}$ and $\phi_j$ are enough to evaluate $P_j$. $M$ should be larger than the largest possible value of an individual cost $M\geq \sup_{\mathbf{Q}}J_j(\mathbf{Q})$. $P_j$ is an aggregate metric used to evaluate both task satisfaction and accumulated cost.

Our objective is to find an optimal group trajectory $\mathbf{Q}^*$, that utilizes the heterogeneous capabilities of the agents to maximize the agent performance and abide by all motion constraints. 
We solve for optimal trajectories as follows : 
\begin{align}\label{Eqn:Central-Objective}
\mathbf{Q}^*=&\argmax_{\mathbf{Q}}\;\sum_{j\in \{1,...,N_A\}}P_j(\mathbf{Q},\phi_j),\\
    &s.t. \;q_j(k+1)\in\;adj(q_j(k)),\notag\\
    &\hspace{20pt}\forall j\in \{1,...,N_A\},k\in [0,...,T_p-1].\notag\notag
\end{align}


We maximize agent performance rather than constraining the system to satisfy all individual specifications. We do so to allow as many agents as possible to satisfy their specifications when not all specifications are feasible. 

\begin{remark}
    For simplicity, we assume the collaboration between agents does not affect the dynamics of the agents which are represented by the constraints in \eqref{Eqn:Central-Objective}. The relaxation of this assumption will be investigated in future work.
\end{remark}


\begin{figure}
    \centering
    \includegraphics[width=200pt]{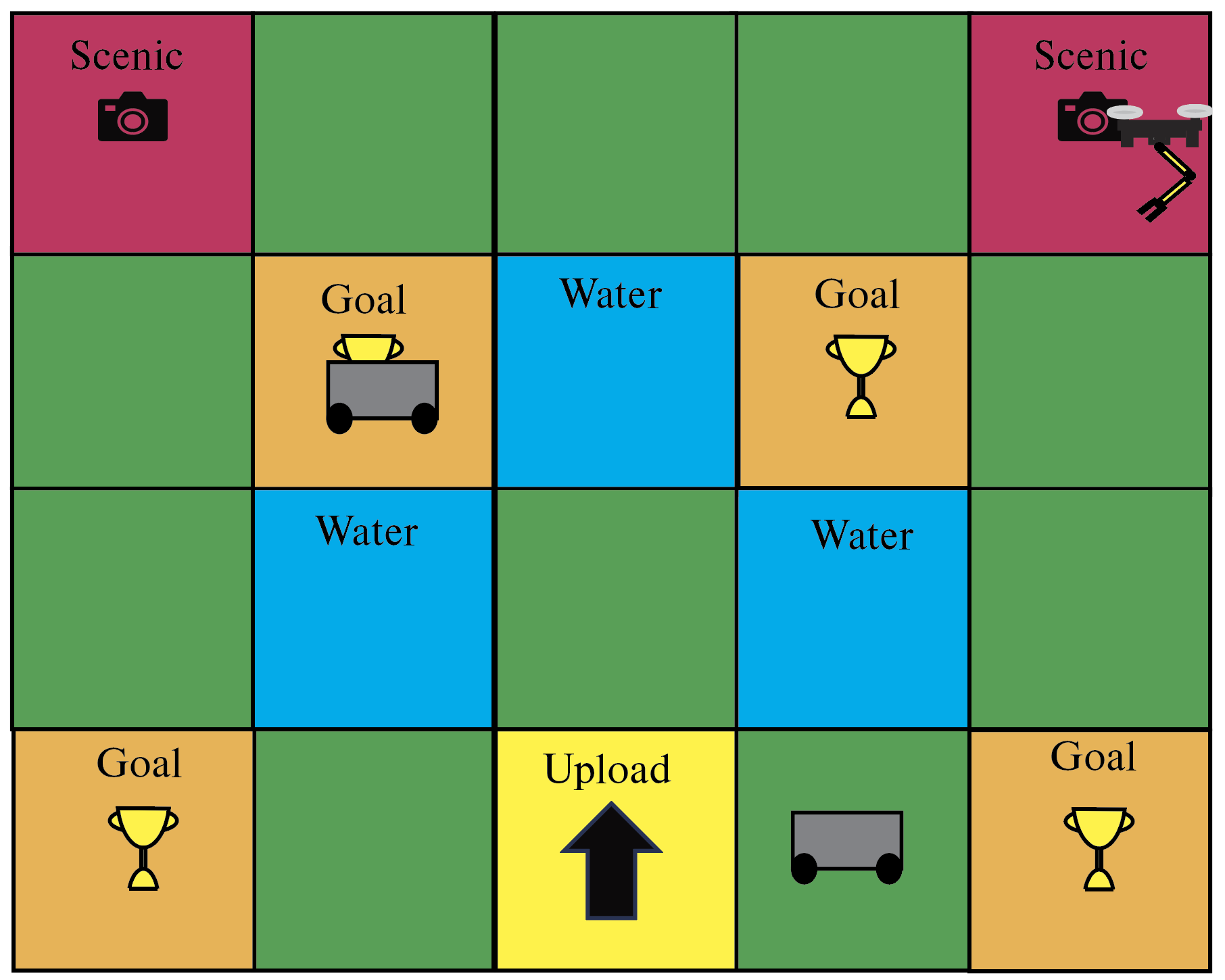}
    \caption{Example: An environment with two ground agents and an aerial agent. The aerial agent must take a picture in ``Scenic" nodes and upload it in ``Upload" nodes. The ground agent must reach ``Goal" and avoid ``Water". In this case, the ground robots can also upload pictures, and the aerial robot can carry one ground robot over the water at a time.}
\label{Fig:example-env}
\end{figure}

\emph{Example:} 
Consider a system composed of three agents ($N_A=3$) with one aerial and two ground robots (see Fig. \ref{Fig:example-env}).
The aerial robot must reach a node with the label ``Scenic" to take a picture and upload it to a database at a node with the label ``Upload". The ground robots must reach a region with the label ``Goal" at some point and never touch ``Water". The ground robot can enter the ``Water" regions if it is carried by an aerial agent. The ground robots have a wireless connection to the database that the aerial agent can use to upload its picture when they are together, independently of whether they are in a node with the label ``Upload". The aerial robot can carry one ground robot at a time. 
Agents plan their trajectories using a central computer, but cannot use it to send or receive images. 

We assume that if an agent is in a node where it can perform actions such as taking and uploading pictures, either on its own or in a collaborating way, it does so automatically. All actions happen instantaneously.


The aerial robot has capability $\mathcal{C}_1=\{\text{``carry"}\}$. The ground agents have capabilities $\mathcal{C}_2=\mathcal{C}_3=\{\text{``WiFi"},\text{``wheels"}\}$.
As shown in Fig.~\ref{Fig:example-env}, the environment has labels $\Pi=\{\text{``Upload"},\text{``Scenic"},\text{``Water"},\text{``Goal"}\}$.

We encode that aerial agent can carry the ground agent over water with a CAT $\langle \neg\text{``Water''},\{\text{``carry''},1\},\{\text{``wheels''},1\}\rangle$ defined over the ground robot. Here, with a slight abuse of notation, we use $\neg\pi$ to denote a label that is assigned to all nodes that are not labeled by $\pi$, i.e., $(\neg\pi)\in\mathcal L(q) \iff \pi\not\in\mathcal L(q)$. This CAT allows the aerial agent to assist the ground robot over water if there are no other ground robots in the same node. 
The CAT defined over the aerial robot that encodes the aerial agent's ability to upload pictures using a ground robot is $\langle \text{``Upload''},\{\text{``WiFi''},1\},\{\}\rangle$. These CATs provide a succinct encoding that captures capability-augmenting collaboration. 

The aerial agent's specification is to take a picture at ``Scenic" in the first 6 time-steps and then upload it within 4 time-steps. The ground agents' specification is to visit ``Goal" in the first 10 time-steps and always avoid ``Water" unless carried by an aerial agent. These specifications are encoded as follows:
\begin{align}
    \phi^1 =& \lozenge_{[0,6]}(\langle\text{``Scenic"},\{\},\{\} \rangle\notag\\
    &\land \lozenge_{[0,4]}\langle\text{``Upload"},\{\text{``WiFi"},1\},\{\}\rangle),\label{Eqn:case-study-two-air-spec},
\end{align}
\begin{align}
    \phi^2=\phi^3 =& \lozenge_{[0,10]}\langle\text{``Goal"},\{\},\{\}\rangle\notag\\
    &\land \square_{[0,10]}\langle \neg\text{``Water"},\{\text{``carry"},1\},\{\text{``wheel"},1\}\rangle.\label{Eqn:case-study-two-gnd-spec}
\end{align}

In this section, we defined capability-augmenting collaboration as a method for agents to utilize each other's capabilities to satisfy individual specifications. In the next section, we synthesize a Mixed Integer Program to find optimal group trajectories for the agents.
\section{Approach}\label{Sect:Approach}
\subsection{Overview}
In the previous section, we formulated the problem of synthesizing a group trajectory that satisfies individual MTL specifications using capability-augmenting collaboration as an optimization problem \eqref{Eqn:Central-Objective}.
In this section, we detail our approach for solving \eqref{Eqn:Central-Objective} using a Mixed Integer Program (MIP). To do this, we first need to encode the group trajectories as decision variables and ensure that their motion is constrained as seen in \eqref{Eqn:Central-Objective}. Next, we need to generate variables that allow us to determine the satisfaction of the CATs. Finally, we need to use these variables to solve \eqref{Eqn:Central-Objective}.

\subsection{Mixed Integer Program Formulation}\label{Subsect:MIP}

In this subsection, we define a 
Mixed Integer Program (MIP) that finds a group trajectory for the agents. The discrete environment and binary and integer elements of the CAT encoding make using an MIP necessary for generating optimal group trajectories.

To begin with, we formulate MIP motion constraints for the individual agents. We define binary decision variables for motion as $z_{j,(q,q'),k'}\in \{0,1\},\;\forall j\in\{1,...,N_A\},\;(q,q')\in E, k'\in [0,...,T_{p}]$. If $z_{j,(q,q'),k'}=1$ then, agent $j$ moves from node $q$ to node $q'$ at time $k'$. Otherwise, a different transition is taken. This means that if $z_{j,(q,q'),k}=1$, then $q_j(k)=q'$ and $q'$ is and element of the group trajectory $\mathbf{Q}$. The motion constraints for the trajectory generated at time-step $0$ are as follows:
\begin{align}
&z_{j,(q,q),0} = \beta(q=q_j(0)),\;\;\forall j\in\{1,...,N_A\}, q\in \mathcal{Q},\label{Eqn:MIP-Contraint-1}\\
&\sum_{(q,q')\in E}z_{j,(q,q'),k'}
= 1,\forall j\in\{1,...,N_A\},k'\in[0,...,T_p],\label{Eqn:MIP-Contraint-2}\\
&\sum_{q'\in\{q'|q\in adj(q')\}}z_{j,(q',q),k'}=\sum_{q'\in adj(q)}z_{j,(q,q'),k'+1},\notag\\
&\hspace{30pt}\forall j\in\{1,...,N_A\},k'\in[0,...,T_{p}-1],q\in \mathcal{Q}.\label{Eqn:MIP-Contraint-3}
\end{align}
Constraint \eqref{Eqn:MIP-Contraint-1} encodes an initial node for the agent using the MIP variables. Constraint \eqref{Eqn:MIP-Contraint-2} enforces that an agent may only take one transition at a time. Constraint \eqref{Eqn:MIP-Contraint-3} enforces that agents must either stay in place or move to an adjacent node. These three constraints combined are equivalent to the motion constraint in \eqref{Eqn:Central-Objective}.

We determine $q_j(k)$ for each agent $j$ at each time step $k$, i.e. which nodes are included in $\mathbf{Q}$, as follows where $ind(q)$ retrieves the index of a node $q$:
\begin{align}\label{Eqn:MIP-CAP-Encoding-1}
    ind(q_j(k))= \sum_{q\in \mathcal{Q}}&\Big(\big(\sum_{q'\in\{q'|q\in adj(q')\}}(z_{j,(q',q),k})\big)*ind(q)\Big),\notag\\
    &\forall j \in \{1,...,N_A\},k\in[0,...,T_p].
\end{align}
To evaluate \eqref{Eqn:Central-Objective}, we need to evaluate $\rho$ and $J_j$ in terms of our MIP variables. Evaluating these allows us to solve \eqref{Eqn:Central-Objective} through solving \eqref{eqn:ind-task-perf}. To find $\rho$, we start by encoding the evaluation of CATs into MIP variables. For an individual CAT, we need to retrieve the agents in $I_{c_{aug}}$ and $I_{c_{al}}$ that are in the same node as agent $j$ using \eqref{Eqn:MIP-CAP-Encoding-2}. We assign each node a unique index $ind(q)\in\{1,...,N_Q\}$ where $N_Q$ is the total number of nodes, 
i.e. $N_Q=|\mathcal{Q}|$. Let $\alpha^j_{j',k}$ represent a binary variable that indicates if agent $j'$ is in the same node as agent $j$ at time step $k$. We determine $\alpha^j_{j',k}$ for agents $j'$ with capabilities $c_{aug}$ and $c_{al}$ as follows:
\begin{align}
    &\hspace{10pt}1-\alpha^j_{j',k} \leq  ||ind(q_j(k))-ind(q_{j'}(k))||_1,\label{Eqn:MIP-CAP-Encoding-2}\\
    &\hspace{60pt}\forall j' \in I_{c_{aug}}\backslash j,k\in[0,...,T_p]\notag\\
    &N_Q*(1-\alpha^j_{j',k})\geq||ind(q_j(k))-ind(q_{j'}(k))||_1,\notag\\
    &\hspace{60pt}\forall j' \in I_{c_{aug}}\backslash j,k\in[0,...,T_p],\notag\\
    &\hspace{10pt}1-\alpha^j_{j'',k} \leq  ||ind(q_j(k))-ind(q_{j''}(k))||_1,\notag\\
    &\hspace{60pt}\forall j'' \in I_{c_{al}}\backslash j,k\in[0,...,T_p],\notag\\
    &N_Q*(1-\alpha^j_{j'',k})\geq||ind(q_j(k))-ind(q_{j''}(k))||_1,\notag\\
    &\hspace{60pt}\forall j'' \in I_{c_{al}}\backslash j,k\in[0,...,T_p].\notag
\end{align}
Finally, we retrieve the number of agents with capability $c_{aug}$ and $c_{na}$ in the same node as agent $j$ as follows:
\begin{align}
    &n^{c_{aug}}_j(k) = \sum_{j' \in I_{c_{aug}}\backslash j}  \alpha^j_{j',k},\forall k\in[0,...,T_p],\label{Eqn:MIP-CAP-Encoding-3}\\
    &n^{c_{al}}_j(k) = \sum_{j'' \in I_{c_{al}}\backslash j} \alpha^j_{j'',k},\forall k\in[0,...,T_p].\notag
\end{align}
Next, we define binary variables for the components of the CAT. Let  $z_{j,aug,k}\in \{0,1\}$ be a binary variable that is $1$ if there are at least $m_{aug}$ agents with capability $c_{aug}$in the same node as agent $j$ at time $k$ otherwise it is $0$. Let $z_{j,al,k}\in \{0,1\}$ be a binary variable that is $1$ if less than $m_{al}$ agents with capability $c_{al}$ are in the same node as agent $j$ at time $k$ otherwise it is $0$. We determine the value of these variables as follows:
\begin{align}\label{Eqn:determine-CAT-vars}
    &z_{j,aug,k} \geq \frac{n^{c_{aug}}_j(k)-m_{aug}+1}{|I_{c_{aug}}\backslash j|},\forall k\in[0,...,T_p],\\
    &z_{j,aug,k} \leq \frac{n^{c_{aug}}_j(k)}{m_{aug}},\forall k\in[0,...,T_p],\notag\\
    &1-z_{j,al,k} \geq \frac{n^{c_{al}}_j(k)-m_{al}+1}{|I_{c_{al}}\backslash j|},\forall k\in[0,...,T_p],\notag\\
    &1-z_{j,al,k} \leq \frac{n^{c_{al}}_j(k)}{m_{al}},\forall k\in[0,...,T_p].\notag
\end{align}
Let $z_{j,\pi,k}\in \{0,1\}$ be a binary variable that is $1$ if agent $j$ is in a region with label $\pi$ and $0$ otherwise. Let $\mathcal{L}^{-1}(\pi)$ be the set of nodes with label $\pi$, i.e. $q\in \mathcal{L}^{-1}(\pi)\iff \pi\in \mathcal{L}(q)$. We determine the value of $z_{j,\pi,k}$ as follows: 
\begin{align}\label{Eqn:label-sat}
    &z_{j,\pi,k}=\sum_{q\in\mathcal L^{-1}(\pi)}\;\sum_{q'\in\{q'|q\in adj(q')\}}z_{j,(q',q),k},\forall k\in[0,...T_p].
\end{align}
To quantify the satisfaction of tasks $\phi_j$, we use $\rho(\mathbf{O}_j,\phi_j,k)$. We find $\rho(\mathbf{O}_j,\phi_j,k)$ as a function of our decision variables as follows:
\begin{align}\label{Eqn:Task-evaluation}
    \rho(\mathbf{O}_j, \text{CAT},k)&=2*(\max(z_{j,\pi,k},\\
    &\hspace{10pt}\min(z_{j,aug,k},z_{j,al,k}))-0.5),\notag\\
    \rho(\mathbf{O}_j,\neg \text{CAT},k)&=-\rho(\mathbf{O}_j,\text{CAT},k),\notag\\    \rho(\mathbf{O}_j,\phi_1\land\phi_2,k)&=\min(\rho(\mathbf{O}_j,\phi_1,k),\rho(\mathbf{O}_j,\phi_2,k)),\notag\\
    \rho(\mathbf{O}_j,\phi_1\lor\phi_2,k)&=\max(\rho(\mathbf{O}_j,\phi_1,k),\rho(\mathbf{O}_j,\phi_2,k)),\notag\\ 
    \rho(\mathbf{O}_j,\lozenge_{[k_1,k_2]}\phi,k)&=\max_{k'\in [k+k_1,k+k_2]}\rho(\mathbf{O}_j,\phi,k'),\notag\\ 
    \rho(\mathbf{O}_j,\square_{[k_1,k_2]}\phi,k)&=\min_{k'\in [k+k_1,k+k_2]}\rho(\mathbf{O}_j,\phi,k').\notag\\
    \rho(\mathbf{O}_j,\phi_1 U_{[k_1,k_2]}\phi_2,k)&= \max_{k'\in [k+k_1,k+k_2]}(\min(\rho(\mathbf{O}_j,\phi_2,k'),\notag\\
    &\hspace{50pt}\min_{k''\in [k,k']}\rho(\mathbf{O}_j,\phi_1,k''))).\notag
\end{align}
We can calculate the individual costs, $J_j$, using the binary variables as follows:
\begin{align}\label{Eqn:MIP-Cost-Evaluation}
    J_j(\mathbf{Q})=\sum_{k=0}^{T_p-1}\sum_{q\in\mathcal{Q}}\sum_{q'\in adj(q)\backslash q}z_{j,(q,q'),k}.
\end{align}
Note that this returns the same value as \eqref{Eqn:individual-cost}. The constraints in \eqref{Eqn:MIP-CAP-Encoding-1}-\eqref{Eqn:MIP-CAP-Encoding-3} provide us with infrastructure to determine the satisfaction of individual CATs. We use \eqref{Eqn:MIP-CAP-Encoding-1}-\eqref{Eqn:label-sat} to evaluate the satisfaction of tasks. Using \eqref{Eqn:Task-evaluation} and \eqref{Eqn:MIP-Cost-Evaluation} we can evaluate agent performance as defined in \eqref{eqn:ind-task-perf}. Finally, \eqref{Eqn:MIP-Contraint-1}-\eqref{Eqn:MIP-Contraint-3} provide the motion constraints found in \eqref{Eqn:Central-Objective}. This accounts for all the pieces required to solve for an optimal group trajectory $\mathbf{Q}^*$ as defined in \eqref{Eqn:Central-Objective}.
\subsection{Algorithm}
The algorithm that formalizes a procedure for generating group trajectories for a team of heterogeneous agents is given in Algorithm \ref{Alg:Main-Procedure}

\begin{algorithm}
\caption{Algorithm for Generating Group Trajectories}\label{Alg:Main-Procedure}
\begin{algorithmic}[1]
\State Initialize agents and environment (Section \ref{Subsect:Model-Notation}).
\State Assign agents tasks using MTL (Sections \ref{subsec:MTL} and \ref{subsec:CAT}).
\State Generate binary transition variables ($z_{j,(q,q'),k}$) for each agent at each time step.
\State Generate variables that indicate that agents are in the same node ($\alpha^j_{j',k}$) for each agent.
\State Generate variables used to evaluate the MTL tasks ($z_{j,\pi,k}$, $z_{j,aug,k}$, and $z_{j,al,k}$) for CAT in each agent's individual MTL specification at each time step.
\State Generate MIP constraints (\eqref{Eqn:MIP-Contraint-1}-\eqref{Eqn:label-sat}).
\State Formulate task performance using the binary variables (\eqref{Eqn:Task-evaluation},\eqref{Eqn:MIP-Cost-Evaluation}).
\State Solve MIP to receive optimal group trajectories.
\end{algorithmic}
\end{algorithm}

\section{Comparison with Existing Formulations}\label{sect:expressivity}
In this section, we compare our ability to capture collaboration between agents to previous formulations. We consider the formulations in \cite{Liu2023,Leahy2022,Buyukkocak2021,Cardona2022}. 

We begin with Capability Temporal Logic Plus (CaTL+) as it is a very rich specification language for heterogeneous teams of agents. We consider a disaster response scenario similar to the case study in \cite{Liu2023}.
This example has a set of six agents in a 3x3 grid. Two agents are aerial and four are ground. The agents must pick up and deliver medical supplies to villages affected by an earthquake. The agents initialize across a river from the villages. There is a bridge over the river that must be inspected by an aerial agent before the ground agents can use it. Only two ground robots can use the bridge at a time. In \cite{Liu2023}, the scenario is designed to show the ability to express the need to inspect the bridge before the ground agents use it, a limit on the number of ground agents on the bridge at a time, and a timed delivery task. 

\begin{figure}
    \centering
    \includegraphics[width=150pt]{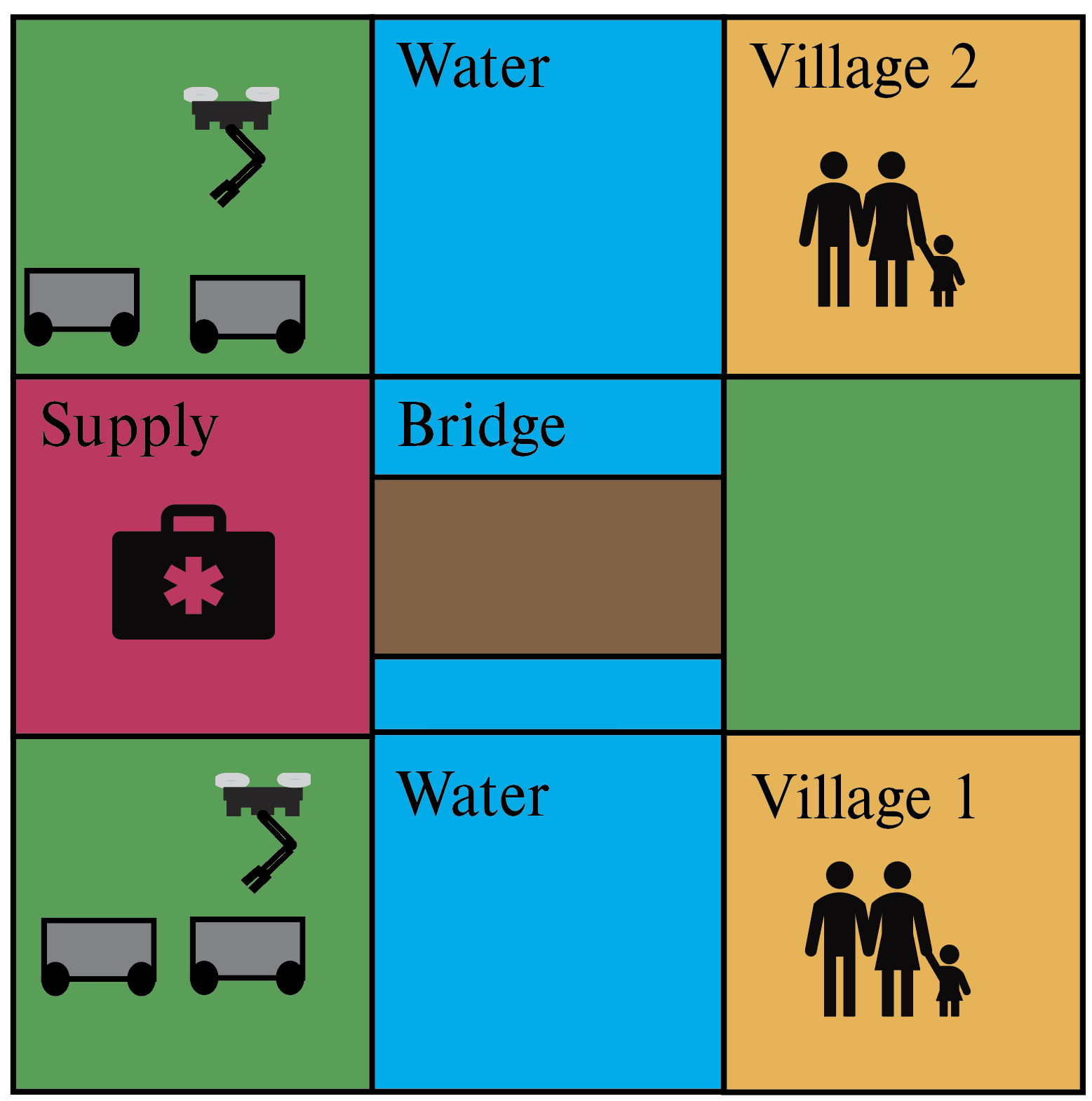}
    \caption{Case Study 1: The agents must pick up supplies in the ``Supply" node and deliver them to ``Village 1" and ``Village 2". Ground robots must avoid ``Water" nodes and cannot visit the ``Bridge" node until it is inspected by an aerial agent. Only two ground agents can visit ``Bridge" at a time}
    \label{fig:Case-study-1}
\end{figure}

In this case, we consider an environment with labels $\Pi=\{\text{``Supply"},\text{``Bridge"},\text{``Water"},\text{``Village 1"},\text{``Village 2"}\}$. The aerial agents have indices $1$ and $2$ and the ground agents have indices $3-6$. The aerial agents have capabilities $\mathcal{C}_j=\{\text{``inspection'',\text{``delivery"}}\},\forall j\in\{1,2\}$. The ground agents have capabilities $\mathcal{C}_j=\{\text{``wheels''},\text{``delivery"}\},\forall j \in \{3,...,6\}$. This is depicted in Figure \ref{fig:Case-study-1}.

In contrast to CaTL+ which creates a single global specification that is assigned to all agents, we specify a task for each agent.
The first aerial agent's specification is to visit ``Village 1" in the first 10 time-steps and not reach ``Village 1" until it acquires supplies at ``Supply". The other aerial agent has roughly the same specification except it must visit ``Village 2". The first two ground agents' specifications are to visit ``Village 1" within the first ten time-steps, not to reach ``Village 1" until it acquires supplies at ``Supply", never to visit ``Water", and to not visit the bridge at the same time as two other agents with capability ``Wheels". The other two ground agents have roughly the same specifications except the agents must visit ``Village 2" instead of ``Village 1". In this case, the CAT encodes the constraint on the maximum number of ground agents on the bridge. Note that we can not capture the constraint that an inspection agent must reach the bridge before the ground agents. This is a constraint that can be captured by CaTL+  by specifying that agents with the capability ``wheels" do not visit ``Bridge" until it is visited by an agent with the capability ``inspection". We cannot capture this because our formulation is focused on collaboration when agents are in the same node. We show our encoding using CATs for the aerial and ground agents respectively as follows:
\begin{align}
 \phi^{1}=&\lozenge_{[0,10]} \langle \text{``Village 1"},\{\},\{\}\rangle\label{Eqn:case-study-one-air-1-spec}\\
& \land(\langle\neg\text{``Village 1"},\{\},\{\}\rangle\notag\\
&U_{[0,10]}\langle\text{``Supply"},\{\},\{\}\rangle),\notag
\end{align}
\begin{align}
 \phi^{3}=\phi^4=&\lozenge_{[0,10]} \langle \text{``Village 1"},\{\},\{\}\rangle\label{Eqn:case-study-one-gnd-1-spec}\\
& \land(\langle\neg\text{``Village 1"},\{\},\{\}\rangle\notag\\
&U_{[0,10]}\langle\text{``Supply"},\{\},\{\}\rangle)\notag\\
 &\land\square_{[0,10]}(\langle\neg\text{``Water"},\{\},\{\}\rangle)\notag\\
 &\land\neg(\langle\text{``Bridge"},\{\},\{\}\rangle\notag\\
 &\land \langle\neg\text{``Bridge"},\{\text{``wheels"},2\},\{\}\rangle).\notag
\end{align}

However, if we allow the aerial agents to carry one ground agent over ``Water" at a time, then this cannot be easily expressed in CaTL+ but can be with our formulation. CaTL+ can specify that multiple agents visit a node labeled ``Water", but directly specify that they meet in the same node. To specify this, we need to specify that they can meet in every state marked ``Water" explicitly. This quickly becomes intractable. In our formulation, the ground agents' tasks become
\begin{align}
 \phi^{3}=\phi^4=&\lozenge_{[0,10]} \langle \text{``Village 1"},\{\},\{\}\rangle\label{Eqn:case-study-one-gnd-2-spec}\\
& \land(\langle\neg\text{``Village 1"},\{\},\{\}\rangle\notag\\
&U_{[0,10]}\langle\text{``Supply"},\{\},\{\}\rangle)\notag\\
 &\land\square_{[0,10]}(\langle\neg\text{``Water"},\{\text{``carry"},1\},\notag\\&\{\text{``wheels"},1\}\rangle)
 \land\neg(\langle\text{``Bridge"},\{\},\{\}\rangle\notag\\
 &\land \langle\neg\text{``Bridge"},\{\text{``wheels"},2\},\{\}\rangle).\notag
\end{align}
This shows a clear example of the differences in expressivity between the two formulations. The specifications given by Capability Temporal Logic (CaTL) \cite{Leahy2022} is a subset of CaTL+. This means that any difficulties experienced in specifying how agents collaborate in CaTL+ are also experienced in CaTL. In CaTL a specified number of agents with a specified capability must visit nodes with a specific atomic proposition (which are similar to labels in our formulation) simultaneously and stay there for a specified amount of time. This makes specifying that agents can satisfy a task without visiting regions with a specific label very difficult.

A similar problem arises in \cite{Buyukkocak2021} and \cite{Cardona2022} where tasks focus on agents with a specified capability to spend a certain amount of time in regions with a specific label or proposition. This, again, makes it difficult to specify how agents can satisfy tasks without reaching nodes with a specific label or proposition. However, these formulations all consider global tasks, meaning it is possible to specify constraints such as the inspection in the example, where one agent must visit a node before a different agent. 

Our formulation provides the tools to specifically capture capability-augmenting collaboration, which is difficult to express in previous formulations. Next, we show the benefit of this collaboration using a characteristic case study.

\section{Case Study}\label{Sect:Case-Studies}
In this section, we use a characteristic case study to demonstrate the effectiveness of the methods defined in the previous sections. We show that the use of CATs improves the individual agent performance of agents at satisfying temporal logic specifications. All simulations were run on a computer with 12 i7-8700K CPUs @ 3.70GHz and 15.5 GB 
of RAM. We used Gurobi as our solver.




\subsection{Case Study, Aerial and Ground System} 
In our case study, we focus on the system with one aerial and two ground agents that we used as an example in the Problem Statement.
We use this case study to demonstrate the benefit of using 
our encoding for capability-augmenting collaboration. Each agent starts in the same node near the environment's center. The aerial agent has an index of $1$ and the ground robots have indices of $2$ and $3$. 

\begin{figure}
    \centering
    \includegraphics[width=240pt]{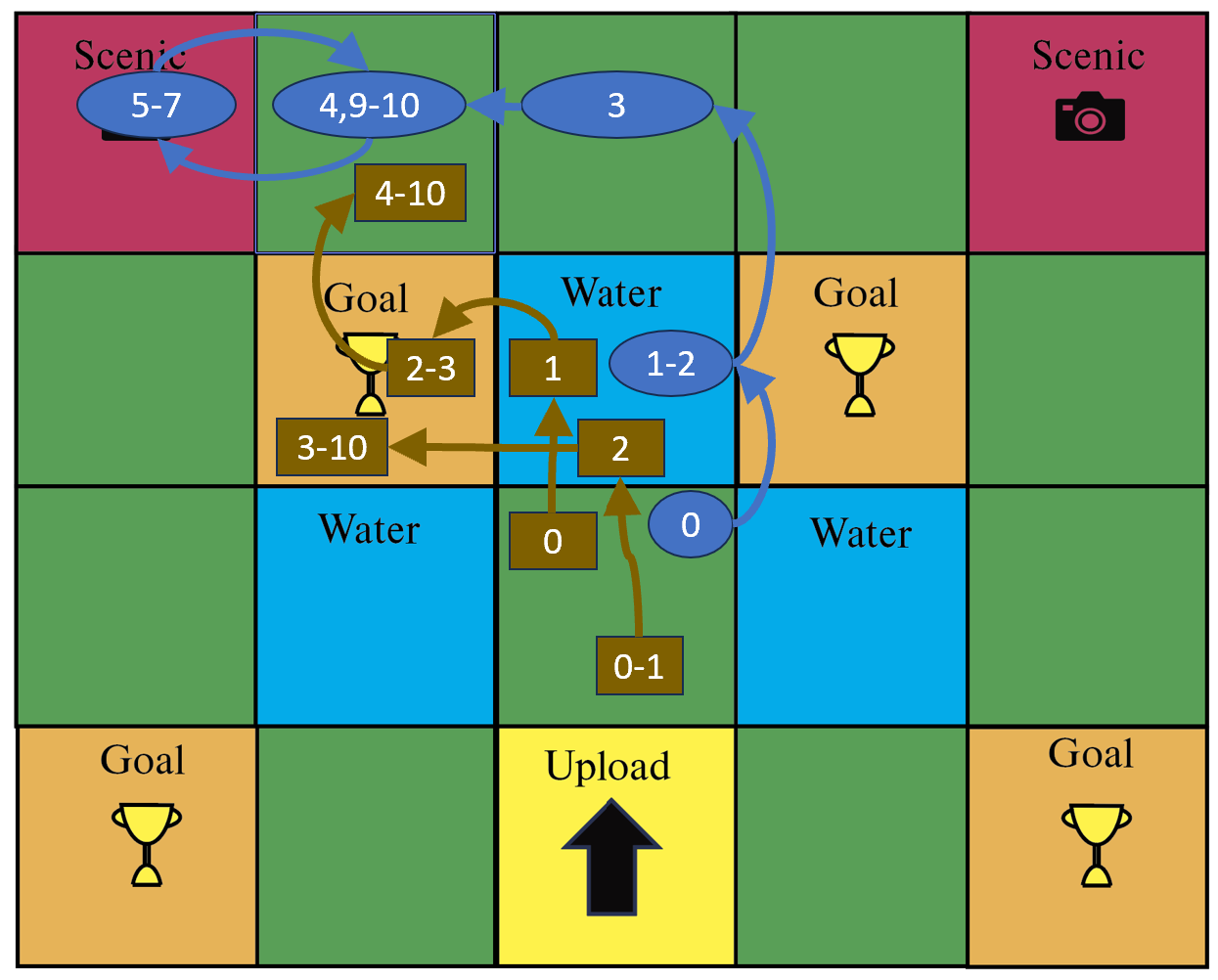}
    \caption{Example solution for the three agent case of case study 2. We show the trajectory for the ground agents' using the rectangles and show the aerial agent's trajectory with ovals. In each rectangle and oval, we specify the exact time steps that the agent occupied the node. We see the aerial agent carry one ground agent over the water at time step 1, then the other at time step 2. The aerial agent takes pictures at time steps 5 -7 and uploads an image using the ground agent at time step 10. The ground robots reach their goal nodes at time steps 2 and 3.}
    \label{fig:case-study-2-traj}
\end{figure}

In this case study, we use an $M$ value of $M=50$ when solving \eqref{Eqn:Central-Objective}. The initial node for each agent is the third row and third column of the environment shown in Figure \ref{Fig:example-env}. For the three-agent case, we found a solution to this problem in an average of $5.43s$ for thirty trials. Each agent completed its task in every trial The average movement for all agents across all trials was $3.33$ transitions. This means that the average individual agent performance was $46.67$. An example solution is shown in Figure \ref{fig:case-study-2-traj}. A solver not using CATs, i.e. task satisfaction is solely determined by the labels of the agents' nodes, could only find solutions for two of the three agents. The motion cost of $2$ meaning the average individual agent performance was $14.67$. The average runtime for the solver not using CATs was $0.41s$. This shows a clear improvement in agent performance when using CATs in this case. However, the runtime using CATs was higher. 

Additionally, we increased the number of agents to 5 with 2 aerial agents and 3 ground agents. For this case, we experienced an average runtime of $67.81s$ over thirty trials. Once again each agent finished its task in every trial when using CATs. We experience an average motion of $3.2$ transitions for each agent across all trials. This means that the average agent performance was $46.80$. For the solver not using CATs, only three of the five agents satisfied their tasks and the average motion cost was $1.8$. This means the average individual agent performance was $8.20$. We experienced improvements in agent performance with the increased number of agents when using CATs.

\begin{figure}
    \centering
    \includegraphics[width=240pt]{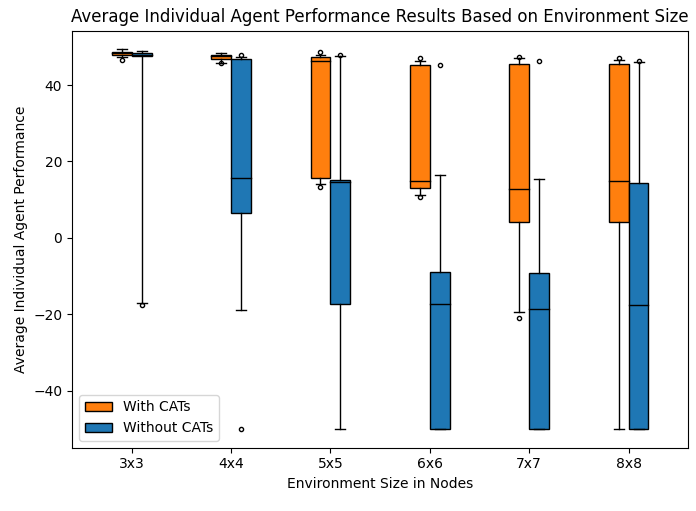}
    \caption{Case Study 2 Results. Each environment size was tested 20 times with three agents in random initial nodes and randomized labels. We randomized labels so that roughly $60\%$ of the nodes were ``Water", $1\%$ of the nodes were ``Upload", $1\%$ of the nodes ``Scenic", and $1\%$ of the nodes were ``Goal". Environments are required to contain at least one of each label. The two formulations were tested in the same random environments. Here, larger agent performance indicates that an agent is better able to satisfy a task. In this case, negative agent performance can only be achieved when the majority of agents in a trial do not satisfy their individual specifications. }
    \label{fig:case-stidy-2-task-perf}
\end{figure}

To further investigate the performance of our formulation, we use the tasks given in \eqref{Eqn:case-study-two-air-spec} and \eqref{Eqn:case-study-two-gnd-spec} and randomize our environment. We generate increasingly large environments to show the effects of CATs as the environment increases in size. We compare this to a system that does not use capability-augmenting collaboration when solving for a trajectory. The results for agent performance are shown in Figure \ref{fig:case-stidy-2-task-perf}. This figure shows the individual agent performance metric defined in \eqref{eqn:ind-task-perf}.
These results demonstrate that the use of CATs consistently results in improved individual agent performance. As the environment becomes larger, the desirable nodes (``Scenic", ``Upload", and ``Goal") become more spread out, making tasks more difficult to satisfy. The use of capability-augmenting collaboration improved the ability of agents to satisfy their tasks in these trials by allowing them to use each other's capabilities when the desirable nodes are harder to reach.


\section{Conclusion}
In this paper, we defined a method to encode capability-augmenting collaboration between agents into the language of MTL by introducing CATs. We formulated a mixed integer program to solve problems using CATs. We demonstrated the capabilities of this encoding to find trajectories using case studies.

In this case, we used CATs in individual tasks to show how heterogeneous agents could leverage each other's capabilities to improve their agent performance. Future work may include creating a global specification formulation that allows individual agents to improve their agent performance using CATs. Additionally, we focus on the problem of generating high-level trajectories for the agents. Future work may focus on creating low-level controllers for agents that utilize these high-level trajectories.
\bibliographystyle{ieeetr}
\bibliography{MainDocument}
\addtolength{\textheight}{-12cm}   




\end{document}